\documentstyle[12pt]{article}
\def\beq{\begin{equation}}
\def\eeq{\end{equation}}
\def\demi{{1\over 2}}
\def\xii{\xi^{({1\over 2})}}

\def\xib{{\overline \xi}}

\def\xij{\xi^{(J)}_M}
\def\xijb{{\overline \xi}^{(J)}_M}
\def\xijbm{{\overline \xi}^{(J)}_{-M}}
\def\xijn{\xi^{(J)}_N}
\def\Jpop#1{M\! \left[J_+\right]_{#1}}
\def\Jmop#1{M\! \left[J_-\right]_{#1}}
\def\Jpmop#1{M\! \left[J_\pm \right]_{#1}}
\def\Jbpmop#1{{\overline M}\! \left[\bar J_\pm \right]_{#1}}
\def\Jtop#1{M\! \! \left[q^{J_3}\right]_{\! #1}}
\def\Jbtop#1{{\overline M}\! \! \left[\bar q^{\bar J_3}\right]_{\! #1}}

\def\Dop#1 #2{M\! \left[D\right]_{#1 #2}}   
\def\Jpo{{M}\! \left[J_+\right]}
\def\Jmo{{M}\! \left[J_-\right]}
\def\Jpmo{{M}\! \left[J_\pm \right]}
\def\Jto{{M}\! \! \left[q^{J_3}\right]}
\def\Jso{{M}\! \! \left[q^{-J_3}\right]} 
\def\ej{e^{-J\alpha_-\Phi}}
\def\mat#1 #2 #3 #4{\Bigl (\begin{array}{cc} #1 & #2 \\
#3 & #4
\end{array}\Bigr )}  

\def\journal#1, #2, #3, #4 { {\sl #1~}{\bf #2~} (#3)  #4 }

\def\cmp{\journal Comm. Math. Phys., }

\def\np{\journal Nucl. Phys., }

\def\pl{\journal Phys. Lett., }

\begin{document}
                                           
\begin{titlepage}

\nopagebreak \begin{flushright}

Imperial/TP/96-97/36\\
hep-th/9703091
 \\
    March 1997
\end{flushright}

\vglue 3  true cm
\begin{center}
{\large \bf
Quantum Group Generators  \\
\medskip
in Conformal Field Theory}\\
\vglue 1.5 true cm
\medskip
{\bf Jens SCHNITTGER}{\footnotesize\footnote{supported in part
by grant ERBCHBICT941380}}\\  
\medskip

\baselineskip .4 true cm    
{\footnotesize Blackett Laboratory, Theory Group, 
Imperial College, \\
London SW7 2BZ, 
United Kingdom\\
and \\
CNRS--Lab. de Math\'ematiques et Physique Th\'eorique (UPRES A 6083), \\
Universit\'e de Tours, 
Parc Grandmont, F-37200 Tours, ~France}    

\end{center}
\vfill
\vglue 1 true cm

\begin{abstract}
Two approaches to the construction of symmetry generators for the
quantum group $U_q(l(2))$ in conformal field theory are presented,
in the concrete context of 2d gravity. The first works with
an extension of the physical phase space and has been successfully
applied already to WZW theory. We show that the result can be used
also for Liouville theory and related models by employing Hamiltonian
reduction. The second is based on a completely new idea and realizes
the quantum group symmetry intrinsically, on the physical phase space
alone.  
\end{abstract}
\vfill
\end{titlepage}

\section{Introduction}

Quantum Groups are known to govern the structure of many integrable
systems, such as Sine-Gordon theory or the XXZ chain, and in particular
of conformal field theory (CFT), where prominent examples are 
Liouville/Toda theory, minimal models, or WZW theory. The quantum group
determines both the operator products and the exchange (braiding) 
relations of the chiral operator algebra. The nonchiral observables
can be viewed as singlets of the quantum group. Thus the quantum group
acts naturally in an extended phase space $\Gamma_L \times \Gamma_R$
 of left- and rightmoving degrees of freedom. This is loosely analogous
 to the case of gauge theories, where observables by definition are
singlets under the gauge group as well, and the extended phase space is 
provided by the gauge fields. The r\^ole of a specified gauge for the
 gauge fields is played by the monodromy of the chiral vertex operators.
 In the example of WZW theory, the general solution $g(x^+, x^-)$ with
 periodic boundary conditions is
$$
g(x^+, x^-)=g_L(x^+)g_R^{-1}(x^-)
$$
in light-cone coordinates $x^\pm=x\pm t$. Here, $g_L$ and $g_R$ are
 periodic up to the monodromy matrix $\gamma$,
\begin{eqnarray}
& g_L(x^++2\pi)=g_L(x^+)\gamma \cr
& g_R(x^-+2\pi)=g_R(x^-)\gamma.
\label{mon}
\end{eqnarray}
Thus, the monodromy matrix $\gamma$ specifies the "gauge", and gauge
 transformations $g_L \rightarrow g_Lg_0, \ g_R \rightarrow
 g_Rg_0$ ($\Rightarrow \ 
\gamma \rightarrow g_0^{-1}\gamma g_0$) do not change the "observable" 
$g(x^+,x^-)$.
 The physical phase space is therefore given by the subset of 
$\Gamma_L \times \Gamma_R$ with equal monodromies ($\gamma_{L}=
\gamma_{R}$),
divided by the set of gauge transformations $g_0$:
\beq
\Gamma_{\hbox{phys}}=\Gamma_L \times \Gamma_R\vert_{(\gamma_{L}=
\gamma_{R})}/G
\label{physphase}
\eeq
where $G$ is the gauge group. The above relation has meaning a priori
only when $G$ is a classical group, and we are considering the classical
phase space. The q-deformation of the classical
symmetry and the quantization of the system need not be
 correlated\footnote{
This is why the name "quantum group" is in general misleading.}; however,
in the case we will consider concretely, that of Liouville  theory,
the two deformations are identified\footnote{The situation is 
somewhat more
complicated in the WZW case \cite{AFSV91}.}.
 The way in which relation Eq. (\ref{physphase})
carries over to the quantum case is then that quantum observables are formed
from left- and rightmoving  vertex operators with equal monodromies in such a 
way that they are invariant under the (suitably defined) action of the universal
enveloping algebra, $U_q(sl(2))$ in the case of Liouville theory. It is important
to note that the monodromy matrix is dynamical; that is, its eigenvalues are 
functionals on the physical phase space, as is explicit in the WZW example above.

\section{Covariant chiral vertex operators   for  $\bf{U_q(sl(2))}$}

In the work of refs. \cite{B88} and \cite{G90} it was revealed that 
2d gravity/Liouville
theory naturally provides a set of chiral vertex operators $\xi(x^+),\xib(x^-)$
with a fixed triangular monodromy, 
which transform covariantly under $U_q(sl(2))$.
They thus form spin $J$ representations $\xi_M^{(J)}$, $\xib_M^{(J)}$.  
Accordingly, the product of chiral vertex operators behaves like 
the product of representations; in particular the interchange of two 
representations
is given by the exchange matrix (or $R$ - matrix):
$$
\xi_{M_1}^{(J_1)}(x_1^+)\xi_{M_2}^{(J_2)}(x_2^+)=
(J_1,J_2)_{M_1 \ M_2}^{M_2' M_1'}
\xi_{M_2'}^{(J_2)}(x_2^+) \xi_{M_1'}^{(J_1)}(x_1^+)
$$
where 
$$
(J,J')_{M\, M'}^{N'\, N}=
\langle J,M\vert \otimes \langle J',M'\vert
{\bf R}
\vert J,N\rangle \otimes \vert J',N'\rangle, 
$$
and
\beq
{\bf R}= e^{(-2ihJ_3 \otimes J_3)}
\left(1+ \sum_{n=1}^\infty \,
{\displaystyle{(1-e^{2ih})^{n}\,e^{ihn(n-1)/2} \over
\lfloor n \rfloor \! !}} 
e^{-ihnJ_3}(J_+)^n \otimes
e^{ihnJ_3}(J_-)^n\right).
\label{exch}
\eeq    
is the universal $R$-matrix of $U_q(sl(2))$. ($\lfloor x\rfloor
:= {q^x-q^{-x}\over q-q^{-1}}$ denotes quantum numbers, as usual.) 
The relation written is
valid for $x_1^+>x_2^+$, and the inverse exchange matrix, with $J_1$
and $J_2$ exchanged, 
is relevant 
in the other case. We normally consider all $\xij$ operators on the 
interval
$x\in [0,2\pi]$, though they are defined for arbitrary $x$. 
The connection between different periodicity intervals is given
 by the 
monodromy operation (cf. section 6). 
Relation Eq.(\ref{exch}) is invariant under the transformation
$\xi_M^{(J)}\to T_{MN}\xi_N^{(J)}$ with $[T^{(J)}_{M_1 M_2},\xi_M^{(J)}]=
0$, provided $T$ fulfills the famous Faddeev-Reshetikhin-Takhtadzhian 
relation
\cite{FRT87}
\beq
T^1T^2 R=RT^2T^1
\label{FRT}
\eeq
On the lefthand side, $T^1$ acts on the first lower index of $R$, and similarly
for the rest. The matrix elements of $T$ can be viewed as elements of
$F_q(SL(2))=U_q^*(sl(2))$, the quantized algebra of functions on the group. 
Their noncommutativity reflects a crucial 
property
of the quantum group symmetry: It is a symmetry of the Poisson-Lie type.
Classically, this means that the Poisson structure of the theory is 
invariant under 
the symmetry only if the symmetry group is equipped with a 
nontrivial Poisson
structure itself \cite{FG92}, namely  the classical limit 
of Eq.(\ref{FRT}) which is given in terms of the classical $r$-matrix
$ R=1+ih r +{\cal O}(h^2)$. 

\section{Quantum Group invariants}
{}From the $\xij$,$\xijb$, it is easy to form singlets under the quantum
group. In Liouville theory, they represent the observables
$\ej$, the exponentials of the Liouville field \cite{GS94} ($\alpha_-$ is 
the semiclassical screening charge of the Coulomb gas):
$$
\ej=\sum_M(-1)^{J+M}q^{J-M}\xij(x^+)\xijbm(x^-)
$$
\beq
\label{exp}
\eeq
The same formula can be reinterpreted as providing local observables in
general Dotsenko-Fateev type models built from integer powers of screenings
in the Coulomb gas picture (without restriction on the central charge) 
and, by taking a limit to rational values
of the central charge, in minimal models \cite{G93}. The transformation
 $\xi_M^{(J)}\to T_{MN}\xi_N^{(J)}$ of section 2 - the quantum Poisson-Lie
 map - is a map from $F_q(\Gamma)$ to $F_q(\Gamma)\otimes F_q(SL(2))$, 
where $F_q(\Gamma)$ denotes the quantized functions on the phase space. 
By dualization, this can be turned into an action of the algebra
 $U_q(sl(2))$ on $F_q(\Gamma)$: For any element $a_q \in U_q(sl(2))$, 
we have
$$
a_q\triangleright\xij =<a_q,T_{MN}\xi^{(J)}_N>
$$
where $<,>$ is the canonical pairing between $U_q(sl(2))$ and 
$F_q(SL(2))$. For $a_q=J_\pm,J_3$ this gives back the usual action of the
$U_q(sl(2))$ generators on a representation of spin $J$:
$$
J_\pm \vert JM>= \sqrt{\lfloor J\mp M\rfloor \lfloor J\pm M+1\rfloor}
\vert J M\pm 1>, 
$$
$$
J_3 \vert JM>=M\vert JM>.
$$
Moreover, on a product of representations,  the
algebra acts by the coproduct. We take it to be given by
$$
\Delta(J_\pm)=J_\pm\otimes q^{J_3}+
q^{-J_3}\otimes J_\pm,
\qquad\Delta(J_3)=J_3\otimes 1+1\otimes J_3.
$$
For the product of a leftmoving and a right-moving representation -
 which differ essentially by complex conjugation - one should use
the above coproduct with $q$ replaced by $q^{-1}$ \cite{G93}.
It is then immediate to check that $\ej$ is indeed invariant under this
action.

\section{Poisson-Lie generators}

In the previous section, we have introduced the action of the quantum 
group "by hand", that is, by direct linear action. On the other hand, the
gauge theory analogy lets us expect that the theory should actually
furnish Hamiltonian generators - the Noether charges - which generate 
the symmetry by Poisson brackets (or commutators in the quantum case):
\beq
\delta f=\{Q,f\}
\label{hamact}
\eeq
where $f$ is any function on the phase space. However, as we have
already remarked in section 2, the symmetry we consider here is of the
Poisson-Lie type. This means that already on the classical level, the
action of the generators is modified. The appropriate generalization
of Eq.(\ref{hamact}) is \cite{AT93}
\beq
\delta_a f=<a,\{m,f\}m^{-1}>
\label{plact}
\eeq
Here $m$ is an element of the dual group $G^*$, the socalled moment 
map, while $a$ is an element
 of the algebra $A$ of $G$, and $<,>$ is the canonical pairing between
 elements of $A$ and $A^*$. On the quantum level, the Poisson-Lie action
is characterized  by commutation relations of the form
\beq
M(a_q) {\cal O}= \sum (a_q^{(1)}\triangleright {\cal O})\  M(a_q^{(2)})
\label{quplact}
\eeq
where $\Delta (a_q)=\sum a_q^{(1)}\otimes a_q^{(2)}$ is the coproduct
of an element of the quantum universal enveloping algebra, and 
${\cal O}$ is an element of the quantized functions on the phase space,
 i.e., an operator. Furthermore, $M(a_q)$ is a homomorphism from
the universal enveloping algebra to the quantized functions on the 
phase space, the (dualized) quantum moment map. The 
action of $ a_q^{(1)}$ on ${\cal O}$ is just the linear action discussed
 above, so for the case at hand we obtain
\beq
M(J_i)\xij= \xijn (J_i^{(1)})_{NM} M(J_i^{(2)})
\label{quplact1}
\eeq

Before we go into the construction of $m$ and its quantum generalization,
it is important to note the following point: Any moment map $m$ capable
of generating $sl(2))$ transformations of the fields $\xij,\xijb$ 
must necessarily be defined on an extended phase space $\Gamma_{\hbox
{ext}}\supset \Gamma_{\hbox{phys}}$. This is because, for the $\xij,\xijb$
of Eq.(\ref{exch}) with their fixed monodromy, the correspondence between
 $\ej$ and $\xij,\xijb$ is one-to-one. Thus, $\xij$ and $\xijb$ can be
 viewed as functions on $ \Gamma_{\hbox{phys}}$, and therefore must be
invariant under the symmetry! The natural way out of this problem is
to formulate $m$ on the extended phase space 
$\Gamma_L \times \Gamma_R$ instead of 
 $ \Gamma_{\hbox{phys}}$, so that also the eigenvectors, and not just the 
eigenvalues, of the monodromy matrix are considered as dynamical. We will
work out in section 5 on the classical level that indeed in this way 
one can obtain the
desired Poisson-Lie action, by Hamiltonian reduction from the WZW theory.
 
Of course, passing to   $\Gamma_{\hbox{ext}}\supset \Gamma_{\hbox{phys}}$
introduces a redundancy, just as working with dynamical gauge fields
does in gauge theory. Surprisingly, there exists a possibility to avoid 
this redundancy, while still achieving the essential part of our goal.
 The idea is to work with generators that act nontrivially on $\xij,
\xijb$ - hence do not leave $\Gamma_{\hbox{phys}}$ invariant 
\underline{globally}
 - but preserve the invariance of a \underline{subset} of observables, 
i.e. functions on $ \Gamma_{\hbox{phys}}$. This approach is in fact 
suggested by a simple observation on the structure of the $R$ - matrix, 
and will be carried out directly on the quantum level in section 6.

\section{Classical moment map for WZW and Liouville theory in extended
 phase space}
\subsection{The case of SL(2) WZW}

The moment map for the WZW model was given  in ref. \cite{FG92}.        
 We recapitulate very briefly the
main statements for the case of $SL(2)$, partially in order to prepare
the  Hamiltonian reduction to Liouville theory. Let us
write the left-moving WZW group element on $\Gamma_L$  as 
\beq
 g_{L}(x^{+})= \mat 1 v_L(x^+) 0 1 
\mat 1 0 w_L(x^+) 1 
\mat {e^{\Phi_L(x^+)}} 0 0  {e^{-\Phi_L(x^+)}}    
\ g_{0L}
\label{param}
\eeq
where $v_L$ and $w_L$ are periodic, while $\Phi_L(x^++2\pi)=\Phi_L(x^+)
+2\pi p_L$. Furthermore, $ g_{0L} $ is
a constant matrix describing the eigenvectors of the monodromy matrix
$\gamma_L$ of Eq.(\ref{mon}),
$$
 \gamma_L =g_{0L}^{-1}e^{2\pi \tau_L }g_{0L}, \ \ \tau_L =
\mat p_L 0 0 {-p_L}  
$$
Similarly, we write\footnote{In the following, we will mainly concentrate
on the left-movers, as the story for the right-movers is very similar.}
$$
 g_{R}(x^{-})=\mat 1 0 w_R(x^-) 1  
\mat 1 v_R(x^-) 0 1
\mat e^{\Phi_R(x^-)} 0 0  {e^{-\Phi_R(x^-)}} 
\ g_{0R}
$$
with the corresponding properties of $p_R$ and $g_{0R}$. On the physical
phase space $\Gamma_{\hbox{phys}}$ we have $g_{0L}=g_{0R}$ and $p_L=p_R$.
The symplectic form $\Omega$ can be written $\Omega=\Omega_L - \Omega_R$,
with
$$
\Omega_L(g_L)={k\over 4\pi}\int_0^{2\pi}dx^+ tr[(g_L^{-1}dg_L)\wedge
\partial_+(g_L^{-1}dg_L)]
$$
$$
+{k\over 4\pi}tr[g_L^{-1}dg_L(0)\wedge d\gamma_L \gamma_L^{-1}]
-{k\over 4\pi}\rho(\gamma_L)
$$
and the same with $g_L \to g_R$ for $\Omega_R$. The two-form 
$\rho(\gamma)$ is a priori arbitrary as it drops out of $\Omega$.
 However, it is possible to choose $\rho$ such that $d\Omega_L=0$ and
$d\Omega_R=0$ separately on a dense open subset of the respective phase 
spaces $\Gamma_L, \Gamma_R$. Following ref. \cite{FG92}, we take
$$
\rho(\gamma)=tr\{(\gamma^-)^{-1}d\gamma^- \wedge (\gamma^+)^{-1}d\gamma^+
\}
$$
Here $\gamma=\gamma^-(\gamma^+)^{-1}$ is the triangular decomposition
of $\gamma$, i.e. 
$$
\gamma^+=\mat {\kappa}  {*} 0 {1/\kappa} , \  \gamma^-=
\mat {1/\kappa}  {0} {{*}} {\kappa}. 
$$ 
The structure of $\Omega_{L,R}$ admits a
PL symmetry $g_l\to g_L g, \ g_R\to g_R g$ ($g$ a constant matrix), 
provided we define on $F(G)$ the Poisson structure
$$
\{g_1,g_2\}:=[g_1g_2,r]
$$
Here, $g_1:=g\otimes 1, \ g_2:=1\otimes g$, and $r={\pi\over k}(e_+
\otimes e_- - e_-\otimes e_+)$ with $e_\pm$ the raising/lowering 
generators of $sl(2)$. Furthermore, $r$ is related to the solutions
$r^\pm$ of the classical Yang-Baxter equation
for $sl(2)$ via
\beq
r^\pm=r\pm {2\pi\over k}C, \ C=\sum_a t^a\otimes t^a
\label{r}
\eeq
being the quadratic Casimir. 

The generator of this PL symmetry -
 the classical moment map - is just given by the components $(\gamma^+,
\gamma^-) \in G^+\otimes G^- \subset G^*$ of the monodromy matrix. One
can indeed deduce directly from the symplectic forms $\Omega_{L,R}$
that for $m_L=(\gamma^+_L,\gamma^-_L)$,
$$
<\epsilon, \{m_L,g_L\}m_L^{-1}>=-tr_1(\epsilon\otimes g_L r^+)
+tr_1(\epsilon\otimes g_L r^-)=
-{4\pi\over k}\sum_a tr(t^a\epsilon)g_L t^a
$$
where $\epsilon$ is any element of $sl(2)$. This means that $m_L$
 (and similarly $m_R$) properly generates the linear action of the
algebra on $g_L$,
$$
\delta g_L=g_L \epsilon_a t^a, \ \epsilon_a = tr(t^a\epsilon).
$$
\subsection{Hamiltonian reduction to Liouville theory}

Using the above result, we can now proceed to obtain the moment map
for Liouville by Hamiltonian reduction \cite{BFFRW89}. 
We will show that the form of
$m$ remains unaffected by the reduction, so that the same moment map, 
given by the monodromy, 
can be used. The reduction procedure is defined by the constraints
$$
j_L^+=\mu^+, \ \ j_R^-=\mu^-
$$
with
$$
j_L^+=tr(e_+j_L), \ j_L=g_L\partial_+g_L^{-1}
$$
$$
j_R^-=tr(e_-j_R), \ j_R=g_R\partial_-g_R^{-1}
$$
Using the parametrizations Eq.(\ref{param}), this becomes
\beq
\partial_+w_L+2\partial_+\Phi_L w_L=\mu^+, \ \partial_-v_R+2\partial_-
v_R=-\mu^-
\label{const}
\eeq
The Liouville field\footnote{Here $\varphi$ is the classical limit
of $\alpha_-\Phi$ in Eq.(\ref{exp}).} is recovered from the Cartan part of the Gauss 
decomposition of $g=g_Lg_R^{-1}$, 
$$
g=\mat 1 x 0 1 \mat e^{\varphi/2} 0
0 {e^{-\varphi/2}} \mat 1 0 y 0 ,
$$ or
$$
e^\varphi ={e^{2\Phi_L(x^+)}e^{2\Phi_R(x^-)} \over (1-w_Le^{2\Phi_L}
v_Re^{-2\Phi_R})^2}
$$
(note that $g_0$ drops out). Eq.(\ref{const}) implies
$$
w_L(x^+)=\mu^+e^{-2\Phi_L(x^+)}S_L, 
$$
$$
S_L={e^{-2i\pi p}\over 2i \sin 2\pi p}
\int_0^{2\pi}e^{2\Phi_L}+\int_0^{x^+}e^{2\Phi_L}
$$
and similarly for $v_r(x^-)$.\footnote{ If we compare this with the 
general Liouville solution formula, $e^\varphi={A'(x^+)B'(x^-)\over
\mu^2(A-B)^2}$, we see that $A=\mu^+S_L, B^{-1}=-\mu_-S_R$ and $ \mu^2=
\mu_+\mu_-$.} We have the diagonal monodromies $S_L(x^++2\pi)=
e^{2\pi p}S_L(x^+), \ S_R(x^-+2\pi)=e^{2\pi p}S_R(x^-)$. 
In order to study the reduced symplectic structure, let us consider
the pair of dynamical variables
$$
\alpha_L\equiv j_L^+=\partial_+w_L+2\partial_+\Phi_L w_L, 
$$
$$
\beta_L\equiv v_L
$$
and similarly for the right-movers. The constraints $\alpha_L=\mu^+, 
\ \alpha_R=\mu^-$ imply $d\alpha_L=d\alpha_R=0$. One can show, moreover, 
that $\Omega_L$ (or $\Omega_R$) only couples $\alpha_L$ and $\beta_L$
(or $\alpha_R$ and $\beta_R$) with each other, and not with the remaining
variables $\Phi_L, g_{0L}$ (or $\Phi_R, g_{0R}$). They thus form a set
of conjugate variables that becomes decoupled through the reduction;
the $\beta_{L,R}$ are gauge variables that do not enter into $e^\varphi$.
This means in particular that the form of the Poisson brackets between
quantities containing $\alpha$ and $\Phi$ but not $\beta$ (that is, 
functions  on the physical Liouville phase space) is unchanged by the
reduction. Let us now consider the matrix elements
$$
(g_Lg_0^{-1})_{22}=w_Le^{\Phi_L}, \ (g_Lg_0^{-1})_{21}=e^{-\Phi_L}
$$
which are independent of $\beta_L$. After reduction, we rename for
convenience $\psi_i:=(g^{red}_{L}g_0^{-1})_{2i}, \ i=1,2$.  
\footnote{Observe that $A={\psi_1
\over \psi_2}$ (cf. previous footnote).} $\psi_1$ and $\psi_2$ are nothing
but the two solutions of the second order differential equation
$$
\psi_i''+T_{++}\psi_i=0 , \ \ (i=1,2)
$$
($T_{++}=-{1\over 2}(\partial_+\varphi)^2+
\partial_+^2\varphi$ the Liouville
 energy-momentum tensor) which describes the associated linear system
appearing in the Lax pair approach to the theory \cite{B89}, and is closely related
 to the uniformization equation. As the $g_{L2i}$ do not contain 
$\beta_L$, the Poisson brackets for the $\psi_i$ have the same form 
as those for the $(g_Lg_0^{-1})_{2i}$ before reduction, whence\footnote
{This agrees, of course, with the classical limit of the results
 of Gervais and Neveu for the exchange algebra of the quantized
$\psi_i$ \cite{GN84}, if one takes the normalization/notation difference 
$\psi_1= {1\over 2\pi i p}\mu^+ \psi_2^{(\hbox{GN})}, \ \psi_2
=\psi_1^{(\hbox{GN})}$ into account.} 
$$
\{ \psi_i(x^+), \psi_i(y^+)\}={\pi\over 2k}\epsilon(x^+-y^+)
\psi_i(x^+)\psi_i(y^+)  
$$
$(i=1,2)$ and 
$$
\{\psi_1(x^+),\psi_2(y^+)\}=-{\pi\over 2k}\epsilon(x^+-y^+)[\psi_1
(x^+)\psi_2(y^+)
$$
$$
+{4\over e^{-4\pi ip\epsilon(x^+-y^+)}-1}\psi_2(x^+)\psi_1
(y^+)]
$$
where $\epsilon(x)$ is the step function.
{}From the $\psi_i\equiv (g_{L}^{(red)}g_0^{-1})_{2i}$ with diagonal
monodromy we can now reconstruct the general 
$g_{L2i}^{(red)}$, or $\psi_i^{g_0}$, by multiplying with $g_0$: 
 $ \psi_i^{(g_0)}=\psi_j g_{0ji}$,  
and in this way we will obtain the Poisson brackets of the $\psi_i^{g_0}$
on the extended phase space $\Gamma_L$. They are of course just given
 by the known Poisson brackets of the $g_{L2i}$ before reduction, thus
\beq
\{\psi_i^{g_0}(x^+), \psi_j^{g_0}(y^+)\}=\psi_k^{g_0}(x^+) 
\psi_l^{g_0}(y^+)(r^\pm)^{kl}_{ij}
\label{psirel}
\eeq
where the upper sign applies when $x^+>y^+$. 

As the moment map $m=(\gamma^+,\gamma^-)$ does not contain $\beta$
either, we know now that it will continue to work correctly in the
reduced setup. Hence,
$$
\delta_\epsilon \psi_i^{g_0}=<\epsilon,\{m,\psi\}m^{-1}>
$$
is the desired transformation of $\psi_i^{g_0}$ under $sl(2)$ 
transformations, for both L and R sectors. The map
$$
\psi_i^{g_0} \to \psi_j^{g_0}t_{ji}
$$
where $t$ is a solution of the classical limit of Eq.(\ref{FRT}),
$$
\{t^1,t^2\}=[r, t^1 t^2]
$$ 
($r$ as in Eq.(\ref{r}))     
is Poisson-Lie, i.e. leaves the Poisson brackets Eq.(\ref{psirel}) 
invariant. Remarkably, as was established in refs. \cite{B88}, \cite{G90},
 the PL-covariant Poisson brackets Eq.(\ref{psirel}) and their quantum
counterparts Eq.(\ref{exch}) can also be obtained in a smaller phase
space $\hat\Gamma$ where $g_{0L,R}$ is not an independent dynamical 
variable,
but has been "gauge-fixed" to become a function of $p$:
$$
\psi_i^{g^0}\to \psi_i^{\hat g^0}
$$
with $\hat g^0=\hat g^0(p)$. The  $\psi_i^{\hat g^0}$ are different
from the "Bloch waves" \footnote{This name is suggested by their 
diagonal monodromy, i.e. periodic behaviour up to a factor.}
 $\psi_{i}$ where $g_0=1$. If we introduce the notation 
$\xi_{-\demi}^{(\demi)}=\psi_1^{\hat g^0}$ and 
$\xi_{+\demi}^{(\demi)}=\psi_2^{\hat g^0}$, we have (classically)
$$
\xi^{(\demi)}_M= \sqrt{-{\pi\over 2p\sin2\pi p}}\ 
\sum_{j=1}^2\vert \demi p)_M^j \psi_j
$$
with 
\beq
\vert \demi p)_M^1=  
e^{2i\pi M p}, \ \vert \demi p)_M^2=pe^{-2i\pi M p}
\label{psixirel}
\eeq
On the quantum level, the $\xi$ fields are nothing but the $J=\demi$
case of the fields of section 2. 
They fulfill the same relations
Eq.(\ref{psirel}) (or Eq.(\ref{exch}) on the quantum level) 
as the $\psi^{g_0}$ even though they live on a different 
 phase space
$\hat\Gamma$. In fact the $\xi$ fields are functions
on the physical phase space $\Gamma_{\hbox{phys}}$ since they
 have a fixed monodromy, just as the $\psi_i$'s\footnote{Of course, the
monodromy of the $\xi$'s is no longer diagonal; it is upper
triangular \cite{CGS95}.}. Fixing the monodromy
is just a way of dividing out the symmetry group $G=SL(2)$ in
Eq.(\ref{physphase}).  It is clear, therefore, that there cannot
exist a moment map which acts nontrivially on the $\xi$'s while
leaving invariant $\Gamma_{\hbox{phys}}$. However, as we will see in the
next section, it is in fact possible to find a moment map that acts
properly on the $\xi$'s, while leaving at least a large subspace of
$\Gamma_{\hbox{phys}}$ invariant.

\section{Realization of the quantum group
 symmetry on the physical phase space}

We will now turn to a completely new approach \cite{CGS95} \cite{CGS96}
 that tries
to realize the quantum group symmetry on $\Gamma_{\hbox{phys}}$
 rather than its extension $\Gamma_L \times \Gamma_R$. Our starting point 
is Eq.(\ref{quplact}), so  we will work directly on the quantum level.
\subsection{Definition of the action by coproduct}
The basic, and rather surprising 
 observation is that there exists a realization of 
the quantum moment map $M$ in terms of the simplest $\xi$ fields
 themselves, namely $\xi^{(\demi)}_{\pm \demi}$. A particular case
of Eq.(\ref{exch}) is ($x_> > x$)\footnote{We drop the index $+$
on the arguments from now on, as we will discuss only the left movers
explicitly.}
\beq
\xi_{-{1\over 2}}^{({1\over 2})}(x_>) \xi_{M}^{(J)}(x)=
q^M \xi_{M}^{(J)}(x) \xi_{-{1\over 2}}^{({1\over 2})}(x_>)
\label{brxi3+}
\eeq
$$
\xi_{{1\over 2}}^{({1\over 2})}(x_>) \xi_{M}^{(J)}(x)=
q^{-M} \xi_{M}^{(J)}(x) \xi_{{1\over 2}}^{({1\over 2})}(x_>) +
$$ 
\beq
{(1-q^2)\over q^{{1\over 2}}} \langle J, M+1|J_+|J,M\rangle
\xi_{M+1}^{(J)}(x) \xi_{-{1\over 2}}^{({1\over 2})}(x_>).
\label{brxi2+}
\eeq      
This has exactly the form of Eq.(\ref{quplact1}), if we identify, up
to constants,  $\xii_{\demi}(x_>)$ with $M(J_+)$, and
$\xii_{-\demi}(x_>)$ with
$M(q^{J_3})$. The  crucial difference with the general
transformation law is that the role of generators is played by
fields that depend upon the worldsheet variable $x_>$.  
This is possible since  the braiding matrix of
$\xii(x_>)$   with
a general field $\xi_{M}^{(J)}(x)$ only depends upon the sign
of $(x_>-x)$. Thus we may realize the Borel subalgebra ${\cal B}_+$ 
of $U_q(sl(2))$  simply by  the $\xii$ fields taken
at an arbitrary point (within the periodicity interval $[0,2\pi]$) 
such that this difference is positive.   
Accordingly, we will write, keeping in mind the $x_>$
dependence,
\begin{eqnarray}
\Jpop{x_>} \equiv\kappa^{>}_+ \xii_{\demi}(x_>),& \nonumber\\
\Jtop{x_>} \equiv \kappa^{>}_3 \xii_{-\demi}(x_>).&
\label{Jopdef+}
\end{eqnarray}
In order to get agreement with Eq.(\ref{quplact1}), the normalization 
constants $\kappa^{>}_+$ and $\kappa^{>}_3$  need to fulfill    
$
{\kappa^{>}_+\over \kappa^{>}_3}={q^{\demi}\over 1-q^2} 
$. 
A similar logic applies to the other Borel subalgebra ${\cal B}_-$. One
defines
\begin{eqnarray}
\Jmop{x_<} \equiv\kappa^{<}_- \xii_{-\demi}(x_<), &\nonumber\\
\Jtop{x_<} \equiv \kappa^{<}_3 \xii_{ \demi}(x_<), &
\label{Jopdef-}
\end{eqnarray}
with $
{\kappa^{<}_-\over \kappa^{<}_3}={q^{-\demi}\over 1-q^{-2}}
$. 
Here $x_<$ needs to be smaller than $x$, so that the other $R$-matrix
is relevant. 

One can verify now that the action of the $\Jpop{} ,\Jtop{} ,\Jmop{} $
generates the quantum group symmetries of the operator product 
and the braiding
of the $\xij$. Let us demonstrate this for the case of the operator
 product. We consider the product $\xi^{(J_1)}_{M_1}(x_1) 
\xi^{(J_2)}_{M_2}(x_2)$ of two general $\xi$ fields. If $x_> >x_{1,2}$
we have, say,   for $J^a\in {\cal B}_+$,
$$
{M}[J^a]_{x_>} \, \xi_{M_1}^{(J_1)}(x_1)
\xi_{M_2}^{(J_2)}(x_2)=
\xi_{N_1}^{(J_1)}(x_1)
\xi_{N_2}^{(J_2)}(x_2) \Lambda_{bc}^a \times
$$
\beq 
\left\{
\Lambda_{de}^b \left[J^d\right]_{N_1M_1}
\left[J^e\right]_{N_2M_2}\right\} {M}(J^c)_{x_>}.
\label{B+prod}
\eeq   
Here $[J^a]_{NM}$ denotes the representation matrix of the generator
$J^a$ in the spin $J_1$ representation, and $\Lambda^a_{bc}$ is the
coefficient matrix of the coproduct.  On the other hand,  
  in ref. \cite{CGR94}
the complete
fusion of the $\xi$ fields was shown to be given 
(in the coordinates of the
sphere) by
$$
\xi ^{(J_1)}_{M_1}(x_1)\,\xi^{(J_2)}_{M_2}(x_2) =
 \sum _{J_{12}= \vert J_1 - J_2 \vert} ^{J_1+J_2}
g _{J_1J_2}^{J_{12}} 
(J_1,M_1;J_2,M_2\vert J_{12})\times
$$
\beq
\sum _{\{\nu\}} \xi ^{(J_{12},\{\nu\})} _{M_1+M_2}(x_2)
\langle\!\varpi _{J_{12}},\{\nu\} \vert V ^{(J_1)}_{J_2-J_{12}}
(e^{ix_1}-e^{ix_2})
\vert \varpi_{J_2}\! \rangle,
\label{fusxi}
\eeq
where $(J_1,M_1;J_2,M_2\vert J_{12})$ are the $q$-Clebsch-Gordan
coefficients, and $g _{J_1J_2}^{J_{12}}$ are the 
so-called coupling constants,  which
depend on the  spins only. The primary fields $V ^{(J)}_{m}(z)$
whose matrix elements appear on the right-hand side are the 
Bloch wave operators with diagonal monodromy, which are direct
generalizations of the $\psi_i$ fields of the previous section
 ($V^{(\demi)}_{-\demi}=\psi_1,V^{(\demi)}_{+\demi}=\psi_2$.). 
They are linearly related to the
$\xi$ fields, similarly to Eq.(\ref{psixirel}). The multiindex $\{\nu\}$
denotes descendants. If we now apply this expansion to both sides of
Eq.(\ref{B+prod}), we find immediately the relation
$$
\sum_{N_1+N_2=N_{12}}  (J_1,N_1;J_2,N_2\vert J_{12}) 
\Lambda_{de}^b \left[J^d\right]_{N_1M_1}
\left[J^e\right]_{N_2M_2}
=
$$
\beq
(J_1,M_1;J_2,M_2\vert J_{12}) \left[J^b\right]_{N_{12}M_{12}},
\label{rec3j}
\eeq
which is just the standard form of the recurrence relation for the 
(q-)$3j$ symbols. A similar analysis for the case of braiding gives
\beq
(J_1,J_2)_{N_1\, N_2}^{P_2\, P_1} 
\Lambda_{de}^b \left[J^d\right]_{N_1M_1} \left[J^e\right]_{N_2M_2}=
\Lambda_{de}^b \left[J^d\right]_{P_2N_2} \left[J^e\right]_{P_1N_1}
(J_1,J_2)_{M_1\, M_2}^{N_2\, N_1}.
\label{defunvR}
\eeq
which  is just the condition that the universal
$R$ matrix interchanges the two coproducts. 
 The same relations follow
from a consideration of the $M$ operators corresponding to ${\cal B}_-$. 
Thus, the $\xij$ generate themselves the symmetries of their operator
algebra in a kind of bootstrap fashion.

\subsection{The algebra of the $M$ operators}
So far we have just considered the action of the $M[J^a]$ on the
$\xij$ but not their commutation relations with each other. If $M$
is a homomorphism as assumed below Eq.(\ref{quplact}),
 they should of course 
just reproduce the $U_q(sl(2))$ algebra. 
It is in fact known - and we will see below - 
 that the most general 
commutation relations compatible with the $U_q(sl(2))$ 
coproduct which appears in Eq.(\ref{quplact1}) are just those
obtained from the standard ones by certain linear redefinitions
of the generators. It will turn out that the $M$ operators should
be viewed as homomorphic to such redefined generators, and not to the
standard ones. 
Let us explain this in 
some more detail. We will ignore at first the complications introduced
by the position dependence of the $M$ operators, and make up for this 
later. The most general commutation relation for
$\Jpop{}$ and $\Jtop{}$ compatible with Eq.(\ref{quplact1}) 
is
\beq
 q\Jpo \Jto -\Jto \Jpo  = C_+
\label{defcp}
\eeq
where $C_+$ is a central term which commutes with all the
$\xi$'s. Similarly, one has  
\beq
\Jto \Jmo  -q^{-1} \Jmo \Jto = C_- .
\label{defcm}
\eeq           
Finally, by considering the action of 
$\Bigl[ \Jpo ,\Jmo \Bigr] $ on $\xij$, one obtains
$$
[\Jpo, \Jmo]= {(\Jto)^2-(\Jso)^{2})\over q-q^{-1}}\ +
$$
\beq
\left (C_3\Jso +C_+\Jmo +C_-\Jpo\right)\Jso
\label{defct}
\eeq        
where $C_3$ is a third central term\footnote{Of course $C_\pm, C_3$
are not central extensions of $U_q(sl(2))$ (which don't exist!).}. 
One can bring these commutation relations
 into the standard form $[J'_+,J'_-]=\lfloor
2J'_3\rfloor, J'_\pm q^{J'_3}=q^{\mp 1}q^{J'_3}J'_\pm$ 
by means of the redefinitions 
$$
J'_\pm:=\rho(\Jpmo \pm {C_\pm\over 1-q^{\pm 1}}\Jso ), \qquad
q^{\pm J'_3}:=\rho^{\pm 1}  {M}\! \! \left[q^{\pm J_3}\right]
$$
with
$$
\rho^{-4}=(q-q^{-1}) \left(C_+ C_- {{1+q}\over{1-q}} -C_3 \right) +1      
$$
However, in the present field theoretic realization $\rho^{-4}$ will be
given in terms of a bilinear in $\xii$ fields, and $\rho=(\rho^{-4})^{
-1/4}$ possesses no clear meaning. Therefore the above redefinition
is formal, and we are forced to stick with the somewhat nonstandard
realization of the $U_q(sl(2))$ commutation relations.
Let us now consider the concrete realization of the $M$ operators
in terms of the $\xii$ fields, which introduces a position dependence.
In this situation, we have to define what we mean exactly when we
talk about commutation relations of the $M$ operators. For instance, in
the case of ${\cal B}_+$ one finds
\beq
q\Jpop{x_>}\Jtop{x_>'}-\Jtop{x_>}\Jpop{x_>'}
=C_+(x_>,x_>')
\nonumber
\eeq
with $x_> > x_>'$. Notice that the positions $x_>,x_>'$ are not 
interchanged as the ordering must be respected also when the $M$ operators
act on themselves. We call this definition of the commutation relations
 fixed point
(FP) commutator. A similar convention applies to the case
of ${\cal B}_-$.  Explicitly, one computes
\begin{eqnarray}
C_+(x_>,x'_>)&=& \sqrt q \kappa^{>}_+\kappa^{>}_3 \>
\xi^{[\demi , \demi](0)}_0(x_>, x_>'),\nonumber\\
C_-(x_<,x_<')&=& {1\over\sqrt q} \kappa^{<}_-\kappa^{<}_3 \>
\xi^{[\demi , \demi](0)}_0(x_<, x_<'), 
\label{Cdef}
\end{eqnarray}    
where $\xi^{[\demi , \demi](0)}_0(x_>, x_>')$ is simply the singlet
$(\demi, M;\demi, -M|0)\xii_M(x_>)\xii_{-M}(x_>')$. The reason that 
$C_+$ is central with respect to any field $\xij(x)$ with $x<x_>'<x_>$,
 and similarly for $C_-$, 
is simply that the $R$-matrix $(J,0)_{M0}^{0N}$ is trivial.

It remains to define commutation relations between elements of
${\cal B}_+$ and  ${\cal B}_-$. Here we meet the obstacle that 
apparently, we cannot hold fixed the positions in the interchange
as required by the FP prescription, because operators from 
${\cal B}_+$ are not defined at points $x_<$, and similarly for
${\cal B}_-$ at $x_>$. 
Fortunately, the
monodromy operation comes to our rescue here. 
 As already mentioned, the $\xij$ fields  are linearly related to the 
Bloch
wave operators with diagonal monodromy (cf. section 5)\footnote{We do not
give the quantum equivalents of the coefficients $|\demi p)_M^j$ of
Eq.(\ref{psixirel}), but they are known (cf. ref. \cite{G90}).}.
{}From this, it is straightforward to deduce 
\begin{eqnarray}
\xii_{-\demi}(x+2\pi)&=& \xii_{\demi}(x), \nonumber\\
 \xii_{\demi}(x+2\pi)&=&2\sqrt q
\cos(2\pi p) \> \xii_{\demi}(x)-q\xii_{-\demi}(x)
\nonumber\\
\end{eqnarray}
and the inverse relation
\begin{eqnarray}
\xii_{-\demi}(x-2\pi)&=&{2\over \sqrt q}
\cos(2\pi p) \> \xii_{-\demi}(x) -{1\over q}\xii_{\demi}(x)\nonumber\\
 \xii_{\demi}(x-2\pi)&=&\xii_{-\demi}(x).
\label{mono}
\end{eqnarray}
Note that $p$ is an operator which does not commute with the $\xij$.

The expression Eq.(\ref{exch}) for the case $x<x'$ is valid for $x'-x
 \in [0,2\pi]$. 
Thus if we start from $\Jmop{x_<}\xij(x)$ with $x\in [0,2\pi]$,
$x_< <0, x-x_< \in [0,2\pi] $, we can reexpress $\Jmop{x_<}$ 
in terms of an operator
at $x_>:=x_<+2\pi \in [0,2\pi]$. Thus we can define
\beq
 \Jmop{x_>}: = \kappa_-^{>} \xii_{-\demi}(x_> -2\pi)
=\kappa_-^>\left({2\over \sqrt q}
\cos(2\pi p) \> \xii_{-\demi}(x_>) -{1\over q}\xii_{\demi}(x_>)\right)
\label{J-+}
\eeq   
with
$$
{ \kappa_-^{>}\over \kappa_3^{>} } =
{ q^{-\demi} \over 1-q^{-2} }              
$$
We can then also identify the Cartan generators at $x_<$ and $x_>$,  
$\Jtop{x_<}\equiv\kappa_3^<\xii_\demi(x_<)
=\Jtop{x_>}\equiv \kappa_3^>\xii_{-\demi}(x_>)$, using again 
Eqs.(\ref{mono}). 
The explicit expressions for the monodromy of $\xii_{\pm\demi}(\sigma ) $
lead to the following relation between $\Jpmo ,\Jto $ and 
$ \cos (2\pi p)$:
\beq
2 \cos (2\pi p) \Jtop{x_>} = (q -q^{-1})
 (\Jmop{x_>} - \Jpop{x_>} ) .
\label{relacos}
\eeq  
This shows that the zero mode $p$ should be viewed as an element of
the universal enveloping algebra. With the definition of $\Jmop{x^-}$ 
above, we can now write the F.P. commutation relations for
$\Jpo$ and $\Jmo$:
\beq
 \Jpop{x_>}  \Jmop{x'_>} -\Jmop{x_>}
\Jpop{x'_>} =
\Dop {x_>,} {x'_>} +
 {\Jtop{x_>} \Jtop{x'_>}  \over q- q^{-1}} .
\label{J+J-sigma}
\eeq       
Here $\Dop {} {} $ is defined by
\beq
\Dop{x_>,} {x'_>} ={1\over  q-q^{-1}} 2\cos
(2\pi p) C_+(x_>, x'_>).
\label{Ddef}
\eeq   
$\Dop {} {}$ is a realization of the operator $(C_+\Jmo +C_-\Jpo )\Jso$
in Eq.(\ref{defct}), and we have
$$
C_3={1\over q-q^{-1}}, \ \ C_+=-C_-
$$
so that there is no term proportional to $\Jso^2$ in Eq.(\ref{defct}).
Note that the operator $\Jso$ is not realized by any combination of
the $\xii$ and $p$.\footnote{However, it can be realized by $\xi^{(-
\demi)}_\demi$, see ref. \cite{CGS95}. One can use this representation
to verify that Eq.(\ref{Ddef}) is obtained by defining $\Jpmo\Jso$
via a renormalized short-distance product.} Rather, it is only the 
particular combination above which can be constructed. However, the 
algebra of $\Jpmo, \Jto,\Dop{} {}$ is equivalent to that of 
$\Jpmo, \Jto,\Jso$, and commutations of $\Dop{} {}$ with the other operators
can always be reexpressed in terms of $\Jpmo, \Jto,\Dop{} {}$ again.
 Thus we obtain a consistent formulation of the algebra in terms of our
F.P. prescription. 

\subsection{Invariance of the physical phase space}

We now discuss the action of the $M$ operators on the physical phase
space, as announced at the end of section 4. The interesting observables,
or functions on the physical phase space, are given by the Liouville
exponentials of Eq.(\ref{exp}). The canonical variables $\Phi$ and 
$\Pi$ can be reconstructed from them by taking suitable derivatives
 \cite{GS94}. Thus the $e^{-J\alpha_-\Phi}(x^+,x^-)$ ($x^+,x^-$ 
varying) can be viewed as probes of different regions of $\Gamma_{
\hbox{phys}}$. According to the remark at the end of section 3, we expect
the $M$ operators for left and right-moving sectors together to be
given by
$$
\Jpmop{x_>^+,x_>^-}^{(LR)}=\Jpmop{x_>^+}\Jbtop{x_>^-}
+
 \Jtop{x_>^+}\Jbpmop{x_>^-}
$$
and
\beq
\Jtop{x_>^+,x_>^-}^{(LR)}=\Jtop{x_>^+}
{{\overline M}\! \! \left[ \bar q^{-\bar J_3}\right]_{\! x_>^-}}  
\label{lrmop}
\eeq
where the ${\overline M}$ operators are defined exactly like the
$M$ operators, but in terms of the $\xijb$ fields, and $\bar q
\equiv q^{-1}$.\footnote{In fact, in order to realize the operator
$ {\overline M}\! \! \left[ \bar q^{-\bar J_3}\right]_{\! x_>^-} $ one needs
to use the field ${\overline \xi}^{(-\demi)}_{+\demi}$ (cf.  last footnote).} 
Inserting the
definitions, one finds
$$
\Jpop{x_>^+,x_>^-}^{(LR)}=e^{-\demi\alpha_-\Phi}(x_>^+,x_>^-)
$$
$$
\Jmop{x_<^+,x_<^-}^{(LR)}=e^{-\demi\alpha_-\Phi}(x_<^+,x_<^-)
$$
\beq
\Jtop{x_>^+,x_>^-}^{(LR)}=\xii_{-\demi}(x_>^+){\overline\xi }^{(\demi)}
_{-\demi}
(x_>^-)
\label{lrmop1}
\eeq
up to irrelevant constants. 
Thus, the action of the lowering/raising operators is just given by the
simplest Liouville exponential itself! Invariance under the action of, 
say, 
$J_+$ thus just means that $e^{-\demi\alpha_-\Phi}(x_>^+,x_>^-)$ and 
$e^{-J\alpha_-\Phi}(x^+,x^-)$ are relatively local. This is certainly
true for spacelike separations, hence in particular if the conditions
$x_>^+ \, > \, x^+, \ x^-_> \, > \, x^-$ are fulfilled\footnote{
(or in fact, if both of them are violated; but  this case is equivalent
to the previous one because of the periodicity of the Liouville exponentials,
 which allows to apply a double monodromy operation to both $\xii_{\pm\demi}
(x^+)$ 
and
${\overline \xi}^{(\demi)}_{\pm\demi}(x^-)$, leading back to 
the previous case.)}. 
 On the other
hand, if (exactly) one of the conditions is violated, the two exponentials are
not mutually local and thus $e^{-J\alpha_-\Phi}(x^+,x^-)$ is not invariant
under the action of the $\Jpmop{}^{(LR)}$ operator. 
Now clearly we can move the window of
"allowed" $x^\pm$ values anywhere we like by changing $x^+_>, x^-_>$
appropriately, using the monodromy operation if necessary. Thus, the 
present formulation of the quantum group generators allows to consider
  the subspaces of $\Gamma_{\hbox{phys}}$ given by local functionals in a
$2\pi$ neighborhood of the values $x_>^+,x^-_> $, which in turn can
be chosen arbitrarily. In other words, one has movable windows on the
physical phase space which are left invariant by the $M$ operators.  

\section{Conclusions}

We have presented two rather different approaches to the problem
of constructing moment maps - or q-Noether charges - for the
quantum group symmetry in conformal field theory. The first one uses the
standard apparatus of Poisson Lie symmetry,
 and we employed Hamiltonian reduction to obtain the moment map for
Liouville theory from the known result for the WZW case. 
Our analysis was obviously incomplete as we presented only the classical
moment map. However, it is conceptually clear how to extend the analysis
to the quantum case and one expects no major technical obstacles;
work on this is in progress \cite{KS}. 
Furthermore, an extension to general Toda 
theories seems both desirable and feasible. 

On the other hand, in the "intrinsic" approach working on the physical
phase space we were able to obtain a realization directly on the
quantum level. 
 However, this approach, which starts from a 
"phenomenological" observation on the structure of the $R$ matrix, is 
conceptually still much less understood. Its most striking feature
 is perhaps the position dependence of the generators, which suggests
an embedding into a Kac-Moody type structure, though of course Liouville
theory a priori just has a global $SL(2)$ symmetry. It is intriguing that
in this way one can realize a kind of BRST transformations on a 
gauge-fixed theory without extending the
 phase space. Clearly, it is
important to understand better the deeper meaning of this new 
"weak" symmetry of the physical phase space, which leaves freely
movable parts of it invariant but never all of it. We remark that
the construction can be extended to $U_q(sl(2))\otimes U_{\hat q}(sl
(2))$ \cite{CGS95},
 where $\hat q$ is the deformation parameter which corresponds 
to the second screening charge $\alpha_+$. This symmetry is well
known within the description of minimal models \cite{AGS89}, where only
half-integer positive spins are relevant, but much less in the context
of arbitrary spins as required for Liouville theory. 

In ref. \cite{CGS96}, it was shown how the second approach can also be
formulated directly in the Bloch wave ($\psi$) basis, which is
closely related to the standard Coulomb gas picture of conformal
field theories. Remarkably, in this
realization one finds a hidden $U_q(sl(2))\times U_q(sl(2))$ symmetry
\footnote{already in the case of one screening charge !}
with a nonstandard coproduct structure and an additional constraint,
which describes the symmetries of the Bloch wave operator algebra
in the same way as the $M$ operators did for the $\xij$ fields.
  
\paragraph{Acknowledgements:}
The  work contained in section 5 has been done in collaboration
with C. Klimcik, while the results in section 6 
were obtained together with  E. Cremmer and J.-L. Gervais. I would like
to thank all of them for a fruitful collaboration, and B. Jurco and
A. Alekseev for many helpful discussions.


\begin{thebibliography}{99}

\bibitem{AFSV91} A. Alekseev, F. Faddeev, M. Semenov-Tian-Shanskii,
A. Volkov, CERN-TH-5981-91, Jan. 1991. K. Gawedzki, "Quantum Group
Symmetries in Conformal Field Theory", Kyoto 1992 proceedings,
Quantum and Noncommutative Analysis, p. 239; hep-th/9210100. 
\bibitem{B88} O. Babelon,  \pl B215, 1988, 523.
\bibitem{G90} J.-L. Gervais, \cmp 130, 1990, 257.
\bibitem{FRT87} L. Faddeev, N. Reshetikhin, L. Takhtadzhian, 
"Quantization of Lie Groups and Lie Algebras", LOMI-E-87, 1987. 
\bibitem{FG92} F. Falceto, K. Gawedzki, lectures at the 
XXVIIIth Karpacz winter school, Feb. 1992; {\sl 
J. Geom. Phys.} {\bf 11} (1993) 251. 
\bibitem{GS94} J.-L. Gervais, J. Schnittger, \np B431, 1994, 273.
\bibitem{G93} J.-L. Gervais, \np B391, 1993, 287. 
\bibitem{AT93} A. Alekseev, I. Todorov, \np B421, 1993, 413.
\bibitem{BFFRW89} J. Balog, L. Feher, P. Forgacs, 
 L. O'Raifeartaigh,
 A. Wipf, pl  B227, 1989, 214. M. Bershadsky, \cmp
139, 1991, 71. 
\bibitem{B89} O. Babelon, "From Integrable to Conformal Field Theory",
 Lectures given at 20th Int. GIFT seminar, Jaca, Spain, June 1989;
GIFT seminar 1989, p.1 and PAR-LPTHE-89-29, June 1989. 
\bibitem{GN84} J.-L. Gervais, A. Neveu, \np B238, 1984, 125.  
\bibitem{CGS95} E. Cremmer, J.-L. Gervais, J. Schnittger,
\cmp 178, 1996, 147.
\bibitem{KS} C. Klimcik, J. Schnittger, work in progress.   
\bibitem{CGS96} E. Cremmer, J.-L. Gervais, J. Schnittger, "Hidden
$U_q(sl(2)) \otimes U_q(sl(2))$ Quantum Group  Symmetry in 
Two-Dimensional Gravity",  hep-th/9604131. 
\bibitem{CGR94} E. Cremmer, J.-L. Gervais, J.-F. Roussel, \cmp  161,
 1994, 597. 
\bibitem{AGS89} L. Alvarez-Gaume, C. Gomez, G. Sierra, \np B319, 
 1989, 155.
\bibitem{BB92} O. Babelon, D. Bernard, \cmp 149, 1992, 279;
\pl  B260, 1991, 81.

\end{thebibliography}
\end{document}